\documentclass{elsart}
\usepackage{amssymb}
\usepackage{graphicx}
\usepackage{rotating}
\usepackage{tikz}
\usepackage{color}
\begin{document}

\begin{frontmatter}
\title{$^{35,37,39}$S isotopes in $sd-pf$ space : Shell-model Interpretation}
\author{ A. Saxena$^{a}$,}
\author{ P.C. Srivastava$^{a}$,}
%\footnote{Corresponding author Email : pcsrifph@iitr.ac.in}}
\author{J. G. Hirsch$^{b}$,}
\author{V.K. B. Kota$^{c}$,}
\author{M.J. Ermamatov $^{d,e}$}
\address{$^{a}$Department of Physics, Indian Institute of Technology Roorkee 247 667, India}
\address{$^{b}$Instituto de Ciencias Nucleares, UNAM, 04510 M\'exico, D.F., Mexico}
\address{$^{c}$Physical Research Laboratory, Ahmedabad 380 009, India}
\address{$^{d}$Instituto de F\'isica, UFF, 24210-340, Niter\'oi, R. de Janeiro, Brazil}
\address{$^{e}$TPD, National University of Uzbekistan, Tashkent, Uzbekistan}

\date{\today}

\begin{abstract}

The structure of  $^{35,37,39}$S isotopes is described by
 performing comprehensive shell model calculations with  SDPF-U and SDPFMW interactions. Protons and neutrons are restricted to the $sd$-shell
 for $N < 20$,  neutrons start to fill the $pf$-shell for $N > 20$. 
 Natural parity states are described by only in-shell mixing, unnatural parity
 states with 1p-1h inter-shell neutron excitations. With SDPF-U interaction, reported are the results for natural parity states only because this interaction is 
 not suitable for cross shell excitations.
Calculated energy levels, electromagnetic properties and spectroscopic factors are in good agreement with the recently available experimental data. 

\end{abstract}

\end{frontmatter}
% main text

%%%%%%%%%%%%%%%%%%%%%%%%%%%%%%%%%%%%%%%%%%%%%%%%%%
\section {Introduction}
 \label{s_intro} 
 
 The coexistence of normal and intruder (sometimes deformed) configurations in the low energy region~\cite{wimmer10}
 around the shell gaps $N=20$ and $N=28$, and the evidence of the erosion of these gaps~\cite{tohru95}, have been the focus of
 recent experimental investigations on Mg ($Z=12$)~\cite{wimmer10}, Si ($Z=14$)~\cite{bastin07}, S ($Z=16$)~\cite{force10,chevrier12}
 and Ar ($Z=18$) isotopes~\cite{winckle12}.   
 A strongly deformed intruder ground state has been reported for $^{31}$Mg by the Leuven group~\cite{Neyens05,Neyens11}.
 The recent theoretical study of $^{30}$Mg$(t,p)$$^{32}$Mg  reaction  revealed that $0^+$ ground state  wavefunction is dominated by 
intruder (2p2h and 4p4h) configurations up to 95\% level~\cite {Macchi07}. 
 On the other hand, for $^{33}$Mg the parity of the spin $I=3/2$ ground-state remains a puzzle~\cite{Numela01,Tripathi08,Yordanov07,Kanungo10}.
For this region, a reduction in the shell gap for the $s_{1/2}$ and $d_{3/2}$ proton orbitals has
been reported for P, Cl and K isotopes~\cite{gade,sorlin04,rydt10,broda10}.
The particle-hole excitations involving intruder states play an important role in the study of
nuclei lying between the {\it island of inversion} and the valley of stability. 
In Ar, the development of collectivity near the $N=28$ shell gap has been reported in~\cite{sarmi08} where non-axially symmetric 
deformation was assigned to $^{48}$Ar.
In analogy to $^{32}$Mg with $N=20$, a new {\it island of inversion} around $N=40$ 
has been predicted in Ref.~\cite{ljungvall10}. Finally,
the merging of the islands of inversion at $N=20$ and $N=28$, with large scale shell model calculations
using an extension of the so called SDPF-U interaction was reported by Caurier {\it et al.}~\cite{Cau}. The focus in the present article is on S isotopes.

\begin{figure}
\begin{center}
\includegraphics[scale=0.5]{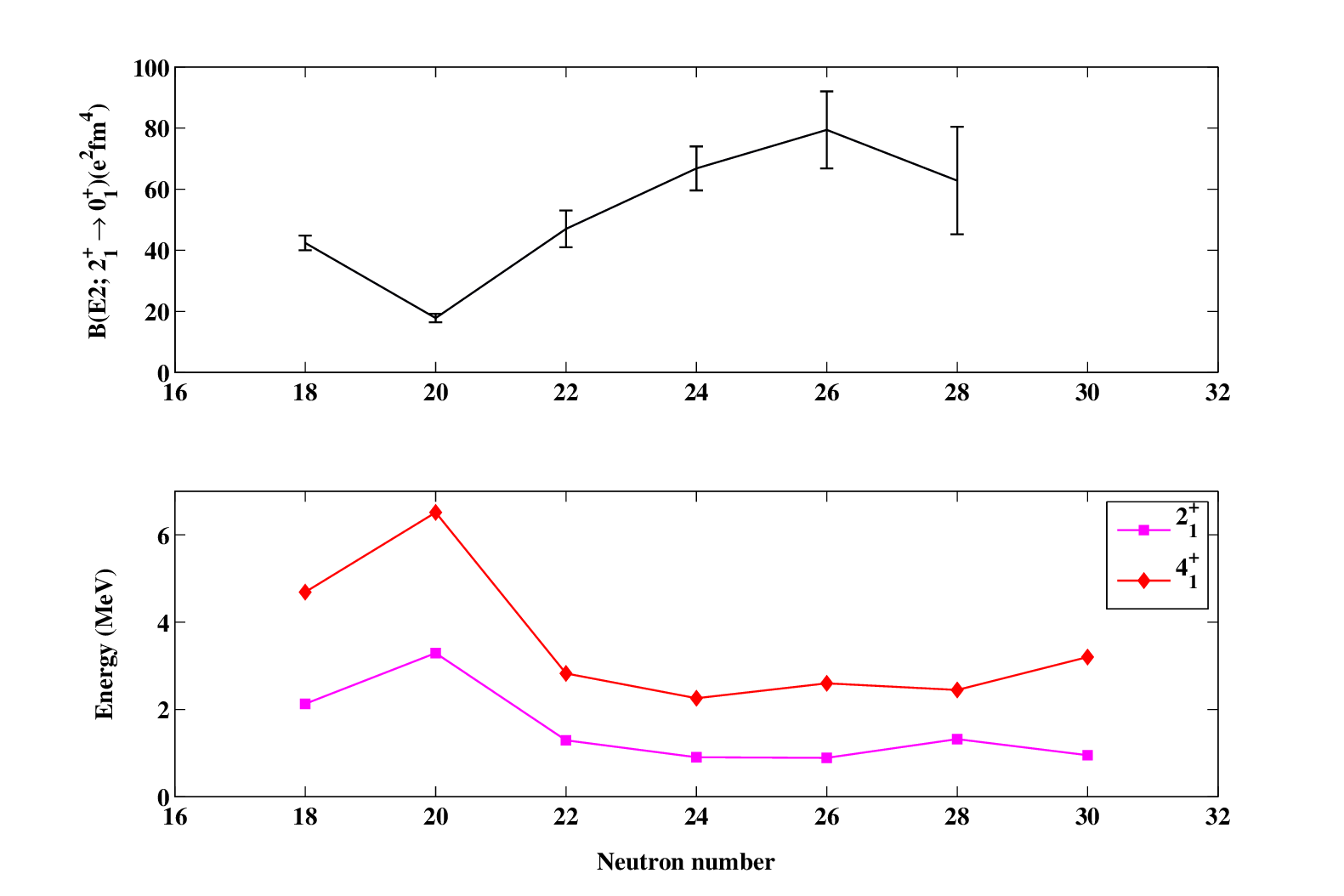}
\end{center}
\label{fig1}
\caption{ Experimental~\cite{nndc} $B(E2)$ values (top) and $2_1^+$ and
 $4_1^+$ energies (bottom) for even sulfur isotopes.}
\end{figure}

Sulfur isotopes exhibit many interesting properties: (i)
$^{40}$S ~\cite{wang10} and $^{42}$S ~\cite{42S} are deformed; (ii) $^{41}$S exhibits $\gamma$-soft properties  and
collectivity~\cite{wang11}; (iii) shape coexistence in $^{43}$S ~\cite{gadefroy09} and shape and configuration
 coexistence in  $^{44}$S is found~\cite{force10,caceres12,tomas11};  (iv) experimental data indicate the erosion of $N=28$ shell gap
in the $^{42}$Si and $^{44}$S  isotones~\cite{bastin07,sohler02}.
With all these, triple configuration
 coexistence has become a topic of current research~\cite{gonzale11}.
Fig.~\ref{fig1} shows the systematics for $B(E2)$ values (top) and the energies of the
$2_1^+$ and $4_1^+$  states (bottom), as a function of the neutron number,
for even-even sulfur isotopes. A clear depletion in the $B(E2)$ values and a peak in the excitation energies are
observed at $^{36}$S with N = 20.  As the neutron number increases, the corresponding $B(E2)$ value also increases, and the excitation energies 
assume nearly constant low values. These two observables are associated with the development of
collectivity. Sulfur isotopes have also an astrophysical importance. In the nucleosynthesis
of heavy Ca-Ti-Cr isotopes, 
neutron-rich sulfur isotopes play an important role~\cite{Davies06}.

From the theoretical side, besides the shell model analysis with realistic interactions presented in~\cite{Aydin,Chapman,Now09},
 Kaneko {\it et al.}~\cite{kaneko11} recently
 reported calculations for the positive parity states in even-even chain of sulfur isotopes using an extended pairing plus quadrupole-quadrupole interaction including
 monopole interactions (EPQQM).

In the present paper, structure of the low energy states in $^{35,37,39}$S isotopes,  with both positive and negative parities, are interpreted in the frame work
of the shell model (SM) in $sd-pf$ space. 
 For proper description of the states with opposite parities which involves cross shell excitations, it is necessary to remove
 the spurious center of mass excitations~\cite{Bouhelal}.
 Present work will add more information to the earlier 
works~\cite{Aydin,Chapman,chap}, as we have performed calculations with the effective interaction SDPFMW that allows for the study of states of both parities. Thus, the present paper gives a more comprehensive SM study for these isotopes. 
The paper is arranged as follows: in Sec.~\ref{sec2}, we discuss the effective interactions 
which have used in this work and also provide some details
of the calculations. 
Results of the calculations for the  energy levels of $^{35,37,39}$S isotopes and also for some of them electromagnetic transition probabilities and spectroscopic factors are 
presented in Sec.~\ref{sec3}. Finally, conclusions are drawn in Sec.~\ref{sec4}.

%=================================================================     
\section{\label{sec2} Effective interactions}

We present SM calculations in the $sd-pf$ space, with
 SDPF-U~\cite{Now09} and SDPFMW~\cite{SDPFMW} interactions using  NuShell~\cite{NuShell} and NuShellX~\cite{NuShellX}.
In the SDPFMW interaction, the $sd$ part is the same as the original USD interaction ~\cite{usd}.
These  interactions are designed to be used in the valence space spanned by the orbitals
$0d_{5/2}$, $1s_{1/2}$, $0d_{3/2}$, $0f_{7/2}$, $1p_{3/2}$, $0f_{5/2}$ and  $1p_{1/2}$,
for both protons and neutrons. 
In the case of SDPF-U, single-particle energies employed
for the  $0d_{5/2}$, $1s_{1/2}$, $0d_{3/2}$, $0f_{7/2}$, $1p_{3/2}$, $0f_{5/2}$ and
$1p_{1/2}$ orbitals are -3.699, -2.915, +1.895, +6.220, +6.314, +11.450, +6.479 MeV,
respectively. For the SDPFMW interaction, the single-particle energies are -3.948, -3.164, +1.647, +14.008, +13.535, +18.524, +13.523 MeV,
respectively.  
Protons and neutrons are restricted to the $sd$-shell for N$<$20,  neutrons start filling the $pf$-shell for N$>$20.
Natural parity states are described with only in-shell mixing, unnatural parity states with 1p-1h inter-shell neutron excitations. 

In general, the 1p-1h state can include large spurious center-of-mass components. To remove this in the present SM
calculations, we have followed the Gloeckner and Lawson approach~\cite{Lawson}, by diagonalizing the modified Hamiltonian.

\begin{equation}
H' = H_{\rm SM} + {\beta_{\rm cm}}H_{\rm cm} 
\end{equation}
\vspace{1mm}
\begin{equation}
= H_{\rm SM} +  {\beta_{\rm cm}} \left\{  \frac { (\sum _{i=1}^A  p_{i})^2 } {2Am} + \frac {1} {2} \frac  {m\omega^2}{A} ( \sum _{i=1}^A  r_{i})^2 -\frac {3} {2}\hbar \omega \right\}, 
\end{equation}
where $H_{\rm SM}$ is the shell model Hamiltonian, $H_{\rm cm}$ is the center of mass operator with $\bf\it{r_{i}}$ and $\bf \it{p_{i}}$ the coordinates and momenta of the
individual nucleons. 
By taking large value of $\beta_{\rm cm}$, $H_{\rm cm}$ contribution to the low-lying states is suppressed (essentially removed). 
In the present work we have taken $\beta_{\rm cm}$ = 10.
 We have also performed calculations for all the three sulfur isotopes with different set of  $\beta_{\rm cm}$  values (5, 8, 10, 12, 15).
Convergence of the energy levels is obtained at $\beta_{\rm cm}$=10.

To obtain natural parity states we have done calculations with SDPF-U and SDPFMW interactions but
 for the unnatural parity states with SDPFMW interaction only because SDPF-U interaction 
 is not suitable for cross shell excitations. This interaction is $0\hbar\omega$ 
 interaction and cannot be used in other circumstances with
some $1\hbar\omega$ and $2\hbar\omega$ matrix elements are missing or zero in this interaction.
For the positive-parity levels in $^{35}{\rm S}$ we have performed 0p-0h calculations using SDPF-U and SDPFMW interactions, i.e. both
protons and neutrons are only allowed to occupy the $sd$ shell. Negative-parity levels can be obtained by exciting one
neutron from the $sd$ shell to the $pf$ shell using  SDPFMW  interaction.  
The parity of the ground state becomes negative from $^{37}{\rm S}$ onwards, since one or more neutrons already occupy the  $pf$ shell.
Thus negative-parity levels are obtained by 0p-0h calculations and 1p-1h calculations give the positive parity bands in case of $^{37,39}{\rm S}$. Now, we will present the SM results. 

%%%%%%%%%%%%%%%%%%%%%%%%%%%%%%%%%%%%%%%%%%%%%%%%%%%%%%%%%%%%%%%%%%%%%%%% 
\begin{figure}
\begin{center}
  \includegraphics[scale=0.6]{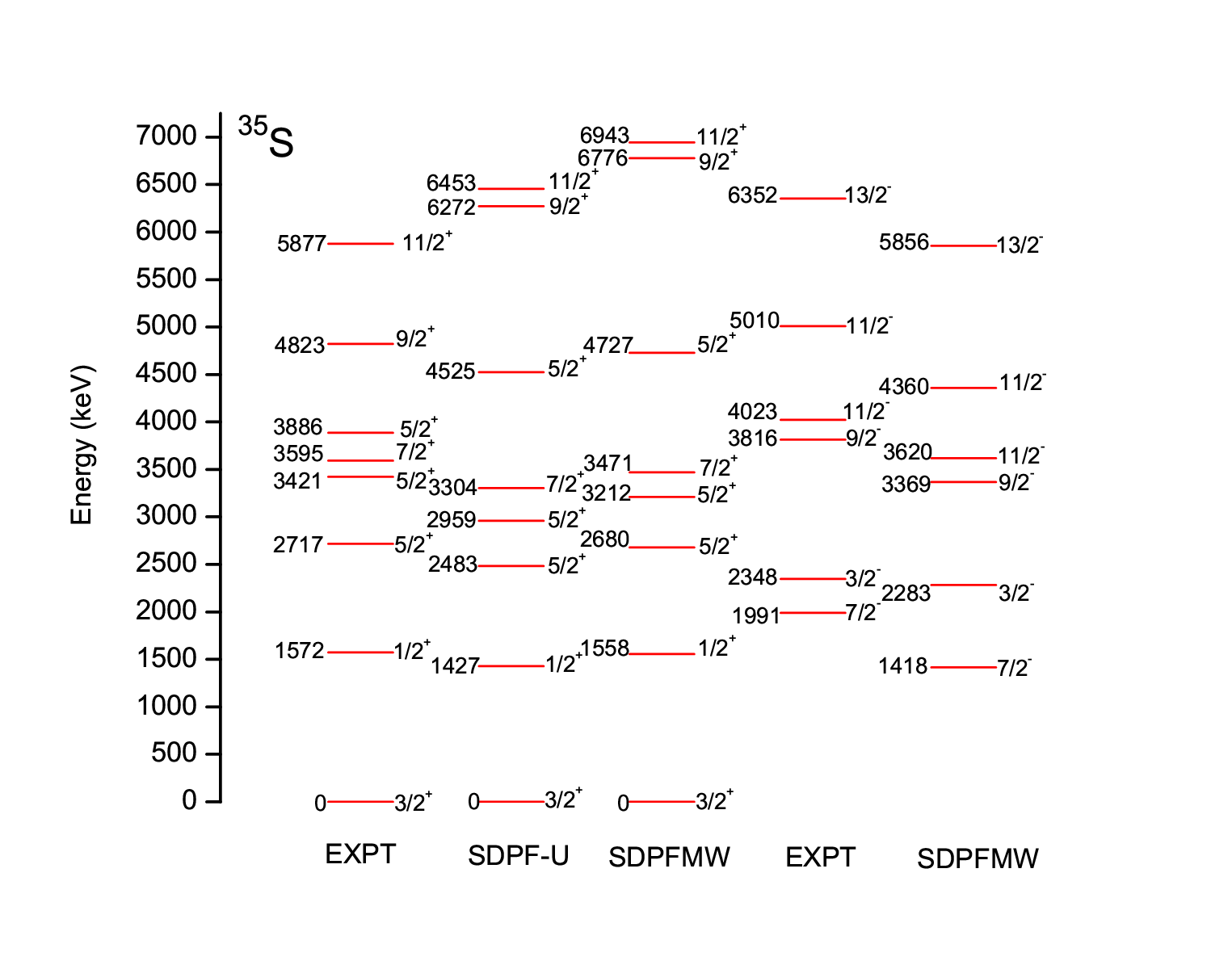}
  \end{center}
\caption{\label{fig2} Calculated and experimental level schemes of $^{35}$S. For the positive parity levels of $^{35}$S the
results with SDPFMW interaction are the same as with USD interaction. }
\end{figure}

%%%%%%%%%%%%%%%%%%%%%%%%%%%%%%%%%%%%%%%%%%%%%%%%%%%%%%%%%%%%%%%%%%%%%%%%
%=================================================================
\section{\label{sec3} Results and Discussion}

\begin{figure}
\begin{center}
  \includegraphics[scale=0.58]{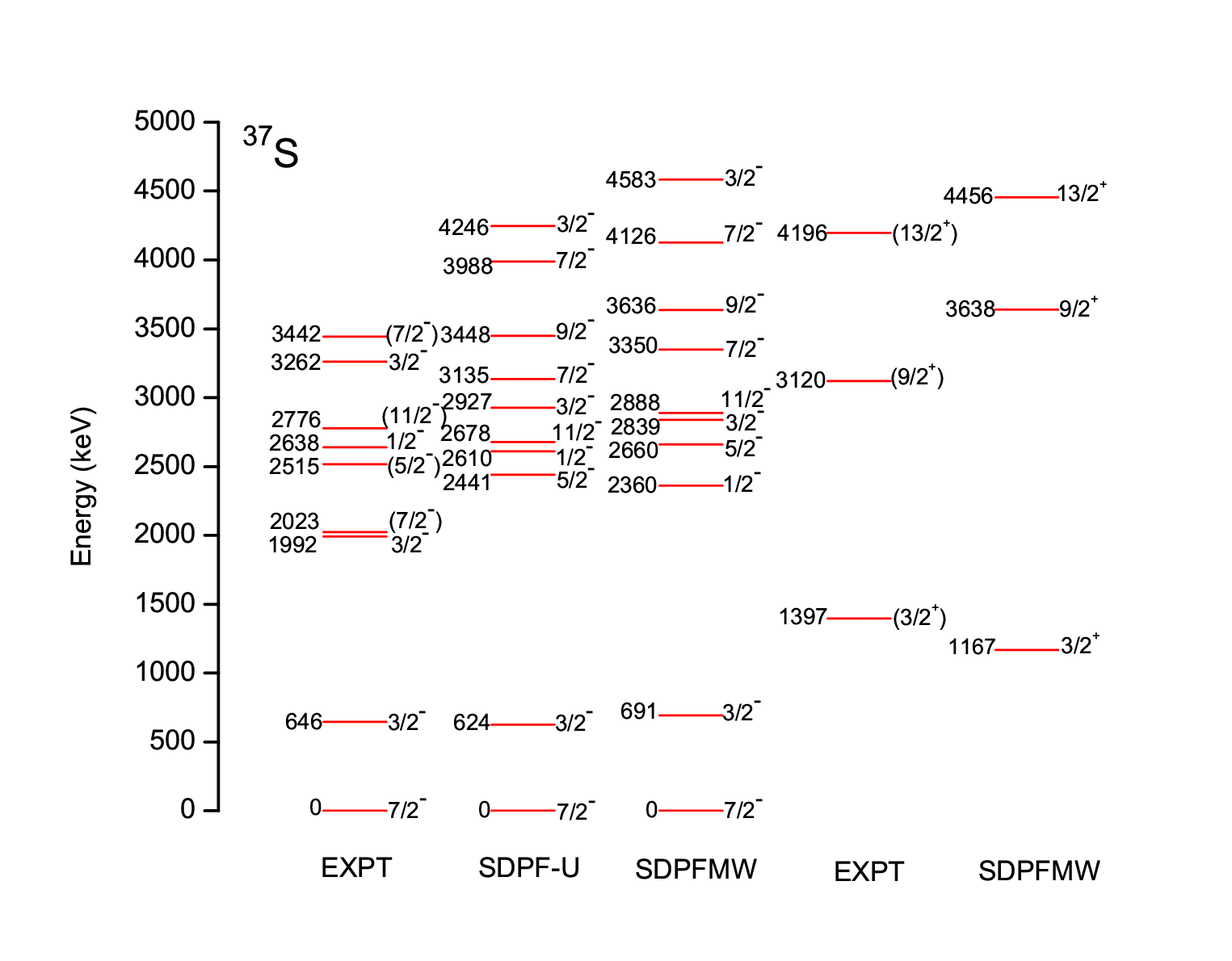}
  \end{center}
  
\caption{\label{fig3} Calculated and experimental level scheme of $^{37}$S.}
\end{figure}

\begin{figure}
\begin{center}
  \includegraphics[scale=0.58]{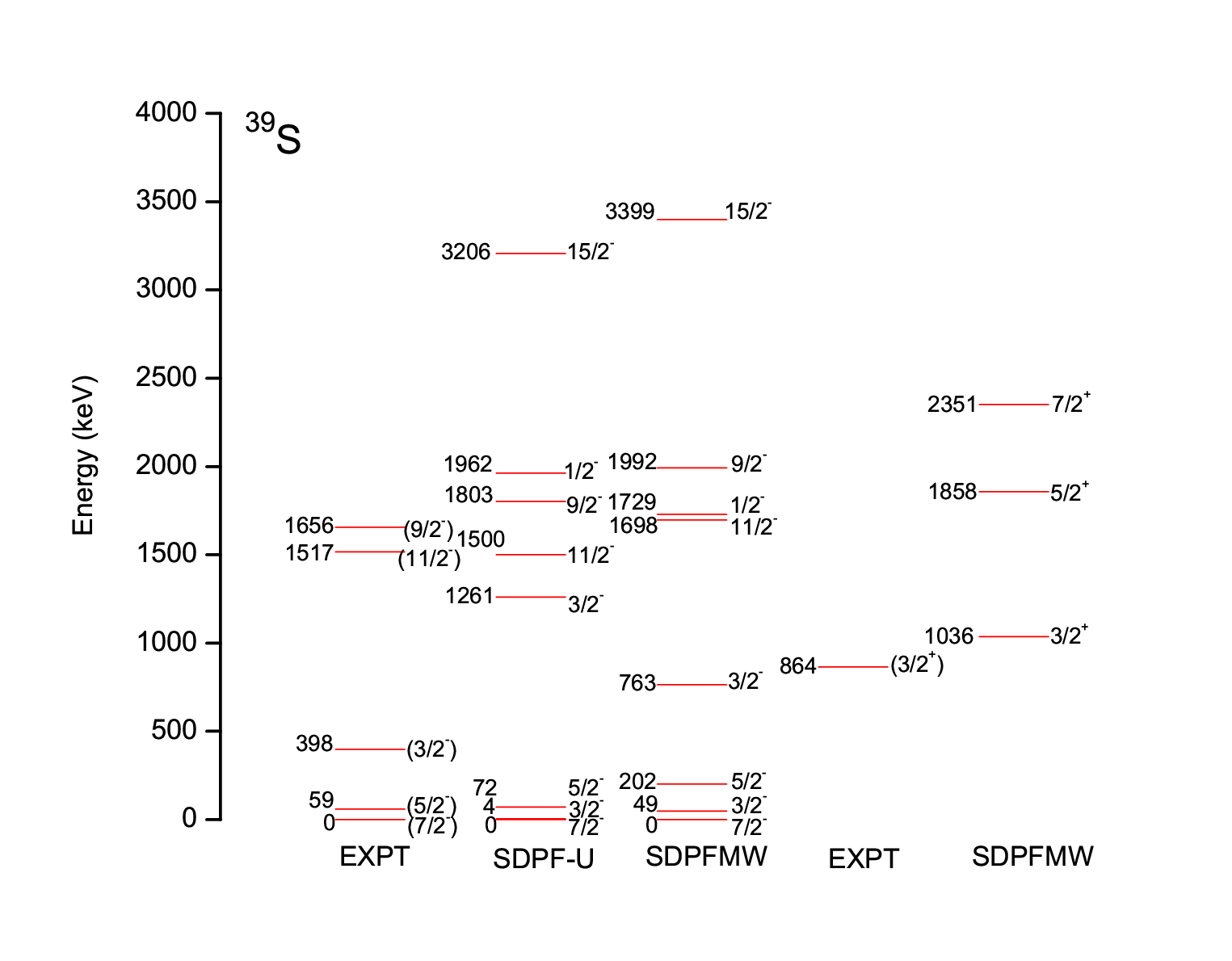}
  \end{center}
\caption{\label{fig4} Calculated and experimental level scheme of $^{39}$S.}
\end{figure}

The comparison of the calculated positive- and negative-parity
states with the experimental data for the odd $^{35,37,39}$S isotopes is presented in Figs.~\ref{fig2}-\ref{fig4}. 
Experimental information for the energy levels in odd sulfur isotopes is available for  $^{35}{\rm S}$,
$^{37}{\rm S}$ and $^{39}{\rm S}$  isotopes in Refs. ~\cite{Aydin,Chapman,chap}. 

The angular momentum of the lowest energy states with both parities can be given a simple qualitative
interpretation, associated with the last uncoupled neutron. In $^{35}{\rm S}$, $N=19$, it occupies the $d_{3/2}$ orbital and corresponding
ground state is $J=3/2^+$.
For $N$ = 21, the $f_{7/2}$ orbital is occupied by one neutron and therefore the ground state is $J=7/2^-$.
For $^{39}{\rm S}$, though with the three neutrons in the $f_{7/2}$ shell, the lowest energy state does not necessarily have J$^\pi$ = $7/2^-$. The three neutron occupancies in $p_{3/2}$
orbital is important as discussed in~\cite{chap}.
As in many other
chains of isotopes, the energy of the excited state with $J=7/2^-$ is high in the lighter isotopes and decreases with increasing
$N$ until it becomes the ground state, while the $J=3/2^+$ becomes the opposite parity excited state with the energy 1397-, and 864-keV for the $^{37}$S and $^{39}$S, respectively. 
In the case of $^{37}{\rm S}$, the SDPF-U interaction gives $J=7/2^-$  as a ground state and $J=3/2^-$ as a first excited state. However, 
these levels are only at a difference of 4 keV in $^{39}{\rm S}$.
One  feature, where the relevance of using the full $pf$  orbitals for the 1p-1h excitations can be seen, is in the energy of the excited opposite-parity states.
The inclusion of the full $pf$ shell decreases the energies of the 
opposite parity states, moving them towards the reported experimental 
values. Now we will present the analysis of the energy spectra of each of the three sulfur isotopes.

%%%%%%%%%%%%%%%%%%%%%%%%%%%%%%%%%%%%%%%%%%%%%%%%%%%%%%%%%%%%%%%%%%%%%%%%%
%%%%%%%%%%%%%%%%%%%%%%%%%%%%%%%%%%%%%%%%%%%%%%%%%%%%%%%%%%%%%%%%%%%%%%%%%%
 
The calculated sequence of positive-parity levels has a clear correspondence with the  experimentally observed ones in $^{35}{\rm S}$ as is shown in Fig.~2. 
The results for 0$\hbar w$ excitation given by SDPFMW interaction are close to the experimental values below $\sim$ 3.5 MeV. 
Negative parity states can be obtained using 1p-1h excitations. 
Previously shell model results for $^{35}$S  use a state-of-the-art PSDPF
interaction~\cite{Bouhelal} are reported in Ref. ~\cite{Aydin}. This interaction is for
psdpf model space with $^{4}$He core. In this, negative parity states are obtained
by allowing one nucleon between major shells.  The PSDPF interaction predicts the experimental
$1/2_1^{+}$ -- $5/2_1^{+}$ -- $5/2_2^{+}$ -- $7/2_1^{+}$ states at 1739, 2679, 3261 and 3544 keV~\cite{Aydin},
while corresponding values with SDPMW interaction are 1558, 2680, 3212 and 3471 keV, respectively.
In the calculation with SDPFMW interaction the $1/2_1^+$ state is very close to the experimental value as compared 
to PSDPF interaction, while PSDPF interaction gives much better agreement with the measured negative parity states ~\cite{Aydin} than with  SDPFMW interaction. 
 
Experimental data are available up to $\sim$ 4.2 MeV for $^{37}{\rm S}$ in Ref.~\cite{Chapman}. Comparison of the
calculated excited state energies with experimental data are shown in Fig.~3.
The ground and first excited states have negative parities, and are correctly reproduced by the calculations.
Experimentally $7/2_2^-$, $7/2_3^-$, $5/2_1^-$, $11/2_1^-$, states are not confirmed. Negative parity states are obtained
by 0p-0h excitation using both interactions and positive parity states by 1p-1h excitations using SDPFMW interaction.
The positive parity states are well reproduced with SDPFMW interaction.
In Ref.~\cite{Chapman}, SM results are given only for the negative-parity levels and they are the same as those shown for SDPF-U results 
shown in Fig.~3. However there the positive-parity levels are 
not given as SDPF-U is not good for them and those in Fig.~3 are the first SM results. 
It is useful to add that for the positive parity states excitation from $pf$ to $0g$ shell
may contribute (also for $^{39}$S discussed below) but the $0g$ orbit is not included in the present study.

For $^{39}{\rm S}$ experimental and calculated results are shown in Fig.~4. 
The SDPF-U calculations predict very good results for these levels.  
The $3/2^-_1$ level is predicted in both calculations a few keV above the $7/2^-$ ground state. 
The SDPFMW calculations predict $3/2^+_1$ level at 1036 keV, while corresponding experimental values is 864 keV.
The order of calculated levels with SDPF-U and SDPFMW interactions are $7/2^-$-$3/2^-$-$5/2^-$, while the experimental one is as $7/2^-$-$5/2^-$-$3/2^-$. 
In the case of $^{39}{\rm S}$, the shell model results
show large configuration mixing, this reflects that the $^{39}{\rm S}$ state is deviating from the single-particle nature.

In case of $^{35}$S, with SDPFMW interaction, the dominant wave function for $7/2_1^{-}$, $3/2_1^{-}$ and
$9/2_1^{-}$ is $\pi(d_{5/2}^6$$d_{3/2}^0$$s_{1/2}^2$)$\otimes$$\nu(d_{5/2}^6$$d_{3/2}^2$$s_{1/2}^2$$f_{7/2}^1$),
$\pi(d_{5/2}^6$$d_{3/2}^0$$s_{1/2}^2$)$\otimes$\\$\nu(d_{5/2}^6$$d_{3/2}^2$$s_{1/2}^2$$p_{3/2}^1$) and
$\pi(d_{5/2}^6$$d_{3/2}^0$$s_{1/2}^2$)$\otimes$$\nu(d_{5/2}^6$$d_{3/2}^2$$s_{1/2}^2$ $f_{7/2}^1$)
with probabilities 36\%, 32\% and 20\%, respectively.
Wave functions  of the natural parity states of $^{37}$S and $^{39}$S are discussed in Refs.~\cite{Chapman,chap}. In the present paper, we have focused on unnatural parity states. 
 In the case of $^{37}$S, unnatural parity states $3/2_1^{+}$, $9/2_1^{+}$ and $13/2_1^{+}$ are coming from the dominant 
 configuration $\pi(d_{5/2}^6$$d_{3/2}^0$$s_{1/2}^2$)$\otimes$ $\nu(d_{5/2}^6$$d_{3/2}^3$$s_{1/2}^2$$f_{7/2}^2$)  
 with  probabilities 43\%, 26\% and 41\%, respectively.  
 In the case of $^{39}$S, the dominant wave function for the unnatural parity  state $3/2_1^{+}$  
 is $\pi(d_{5/2}^6$$d_{3/2}^0$$s_{1/2}^2$)$\otimes$$\nu(d_{5/2}^6$$d_{3/2}^3$$s_{1/2}^2$$f_{7/2}^4$) with probability 17\% and 
 next major one is $\pi(d_{5/2}^6$$d_{3/2}^2$$s_{1/2}^0$)$\otimes$$\nu(d_{5/2}^6$$d_{3/2}^3$$s_{1/2}^2$$f_{7/2}^4$) with probability 16\%. 
 As we move beyond $^{35}$S, there are large number of configurations with very small probabilities. 
 The configurations are strongly mixed.  
 
For $^{37}$S, the SDPFMW interaction predicts $11/2^-$ state at 2888 keV, this state correspond to $^{36}$S$_{2_1^+}\otimes{\nu f_{7/2}^1}$ structure. 
The wave function of $2_1^+$ at 3269 keV
is dominated by the configuration $\pi (d_{5/2}^6s_{1/2}^1d_{3/2}^1) \otimes \nu (d_{5/2}^6s_{1/2}^2d_{3/2}^4)$. The dominant wave function of $^{37}$S$_{11/2_1^-}$  is 
$\pi (d_{5/2}^6s_{1/2}^1d_{3/2}^1) \otimes \nu (d_{5/2}^6s_{1/2}^2d_{3/2}^4f_{7/2}^1$) (87.6\%). This supports the nature of particle-core coupled state
similar to the SDPF-U result reported by Chapman {\it et al.,} in Ref.~\cite{Chapman}.
In the case of $^{39}$S, the SDPFMW interaction predicts $11/2^-$ state at 1698 keV.  The wave function of $2_1^+$ at 1459 keV for $^{38}$S
is dominated by the configuration $\pi (d_{5/2}^6s_{1/2}^1d_{3/2}^1) \otimes \nu (d_{5/2}^6s_{1/2}^2d_{3/2}^4f_{7/2}^2)$. The dominant wave function of $^{39}$S$_{11/2_1^-}$  is 
$\pi (d_{5/2}^6s_{1/2}^0d_{3/2}^2) \otimes \nu (d_{5/2}^6s_{1/2}^2d_{3/2}^4f_{7/2}^3)$ (24.75\%). 
For the $^{39}$S the $11/2^-$ state also favours the particle-core nature but this is not pure particle-core nature as compared to to $^{37}$S because 
of small component of the wave function.

In the Fig.~5, the decomposition of proton and neutron wave functions decomposition is shown. For $^{37}$S, the $11/2^-$ state 
corresponds to $I_p=2^+$$\otimes$$I_n=7/2^-$ configuration with probability of 97\% of the total wave function. As we move to $^{39}$S, 
the $11/2^-$ state shows many components:
$I_p=2^+$$\otimes$$I_n=7/2^-$ (30\%); $I_p=2^+$$\otimes$$I_n=9/2^-$ (2.5\%); 
$I_p=2^+$$\otimes$$I_n=11/2^-$ (7\%): $I_p=2^+$$\otimes$$I_n=15/2^-$ (3\%), 
while the $I_p=0^+$$\otimes$$I_n=11/2^-$ corresponds to the largest component of the wave function (53\%). Thus, $11/2^-$ state of $^{39}$S shows more 
mixed type of configuration. This reflects weak particle-core type of structure.

\begin{figure}
\includegraphics[scale=0.5]{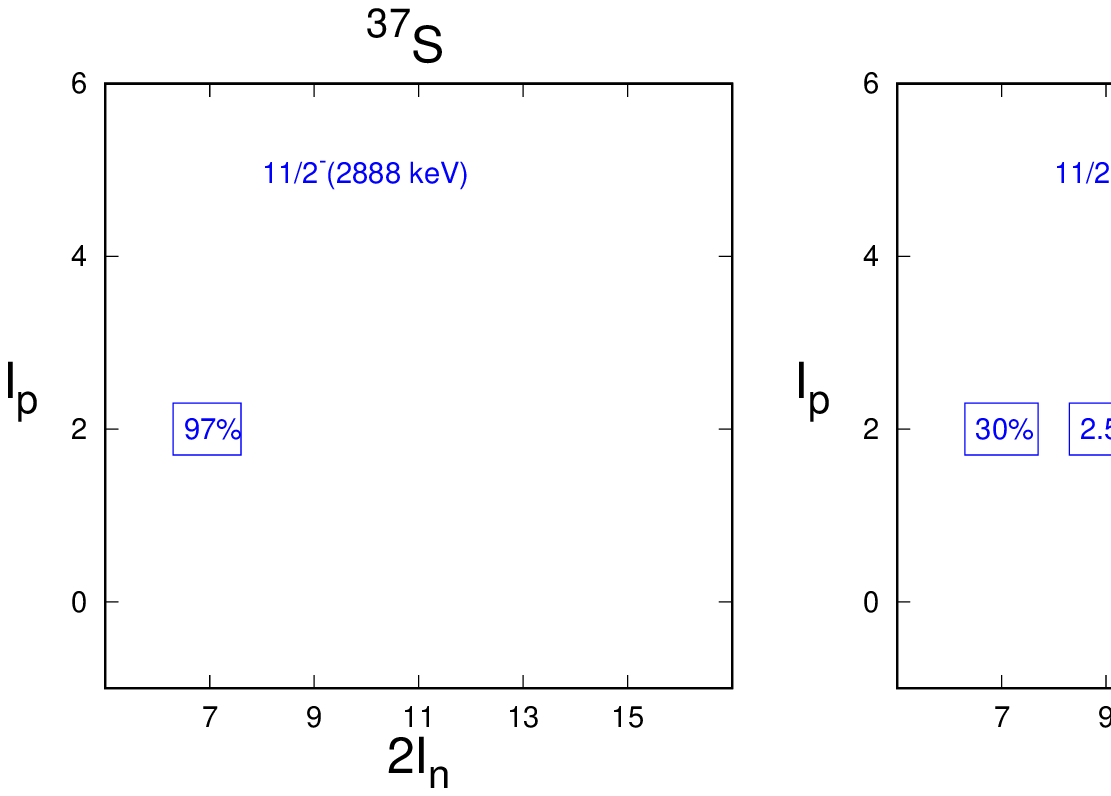}
\caption{\label{fig 5} Decomposition of the total angular momentum of $11/2^-$ state of $^{37}$S and $^{39}$S into their $I_n$ $\otimes$ $I_p$ components.}
\end{figure}

For further understanding the structure of the levels, in table 1, we present the $M1$, $E1$, $M2$ and $E3$ reduced transition probabilities with SDPFMW effective interaction. 
 In this table we have compared our calculated results with PSDPF interaction (Ref.~\cite{Aydin}) also.
The predictions of the SDPFMW interaction are in moderate agreement with $B(M1)$ and $B(E1)$ values.
The $B(M2; 7/2_1^- \rightarrow  3/2_1^+ )$
is 3.4 $\mu_N^2fm^2$ with  $g_s^{\rm eff}$ = $g_s^{\rm free}$. Using the new g factor values $g_{\nu s}^{\rm eff}$ = -2.869, $g_{\nu l}^{\rm eff}$ = -0.1, $g_{\pi s}^{\rm eff}$ = 4.189,
$g_{\pi l}^{\rm eff}$ = 1.1 as suggested in Ref.~\cite{Aydin}, the $B(M2)$ value for this transition with SDPFMW interaction is 1.7 $\mu_N^2fm^2$, corresponding to
the experimental value 1.6 $\mu_N^2fm^2$.
The $B(E3: 7/2_1^- \rightarrow 3/2_1^+)$ is 26.64 $e^2fm^6$ with e$_{\rm eff}^\pi$=1.5$e$, e$_{\rm eff}^\nu$=0.5$e$.
The recent PSDPF interaction gives a much better value of $B(E3: 7/2^- \rightarrow 3/2^+)$ transition in $^{35}$S than does a calculation based on the SDPFMW interaction.
It is possible to get closer value by taking higher value of effective charges.

\begin{table}
\centering 
\caption{ Comparison of calculated and experimental values of $B(M1)$, $B(E1)$, $B(M2)$ and $B(E3)$ transition rates for
 $^{35}$S isotope with effective charges e$_{\rm eff}^\pi$=1.5$e$, e$_{\rm eff}^\nu$=0.5$e$ and $g_s^{\rm eff}$ = $g_s^{\rm free}$ in $\mu_{N}^2$, $e^2fm^2$, $\mu_N^2fm^2$ and $e^2fm^6$ units respectively. $E_\gamma$ is given in keV.}
 %\begin{ruledtabular}
\begin{tabular}{lllllll}
\hline
%&&&\multicolumn{2}{c}{$B(E2)$ (W.u.)}\\
Nucleus & $I_i^\pi \rightarrow I_f^\pi $ & $E_\gamma$  & Expt.  & SDPFMW & PSDPF~\cite{Aydin}\\
\hline
$B(M1)$ &                                 &          &         &       & \\
$^{35}$S      & $1/2_1^+$ $\rightarrow$  $3/2_1^+$   & 1572    & 0.004(1)  & 0.0407 & 0.020\\  
              & $5/2_1^+$ $\rightarrow$  $3/2_1^+$   & 2717    & 0.028(10)  &  0.0738   & 0.038 \\
\hline
$B(E1)$ &                                    &  & &   &  &  \\
$^{35}$S      & $3/2_1^-$ $\rightarrow$  $1/2_1^+$   & 775   & 32(6)$\times$10$^{-5}$  &   14.95$\times$10$^{-5}$  &  54$\times$10$^{-5}$ \\  
              & $3/2_1^-$ $\rightarrow$  $3/2_1^+$   & 2348    & 31(6)$\times$10$^{-6}$  &  24.63$\times$10$^{-5}$   & 10$\times$10$^{-7}$\\ 
\hline
$B(M2)$                                   &  &  & &  &     \\
$^{35}$S      & $7/2_1^-$ $\rightarrow$  $3/2_1^+$   & 1991   & 1.6(5)  &  3.4     & 2.11\\  
              & $3/2_1^-$ $\rightarrow$  $3/2_1^+$   & 2348    & 45(18) &  0.0009  & 0.0044 \\  
\hline
$B(E3)$ &  &  & &   &  \\
$^{35}$S      & $7/2_1^-$ $\rightarrow$  $3/2_1^+$   & 1991   & 115(86)  &   26.64  & 119\\  

  \hline
\end{tabular}
%\end{ruledtabular}
\end{table}

\begin{table}
%\begin{center} 
\caption{ Comparison of calculated and experimental value of quadrupole moments (with $e_p$=1.5e, $e_n$=0.5e ) and magnetic moments ($g_s^{\rm eff}$ = $g_s^{\rm free}$).}
%\begin{ruledtabular}
\begin{tabular}{lllllll}
\hline
    &     & $Q(eb)$  & &  & $\mu $($\mu_N$) \\ \hline
$J^\pi$ & Expt. & SDPF-U  &  SDPFMW  & Expt. & SDPF-U & SDPFMW \\ 
\hline
$^{35}{\rm S}$   &  &&      & &    &         \\
$3/2_1^+$ & +0.0471(9)  &+0.055 &+0.053  &(+)1.00(4)& +1.084 & +1.060\\

$5/2_1^+$ & N/A & -0.0038 & -0.0067 & N/A   &+1.787 & +2.020 \\

\hline
$^{37}{\rm S}$   &        & &    &         \\
$3/2_1^-$  & N/A&-0.041 &  -0.039  & N/A&  -2.002 &  -1.900\\

$5/2_1^-$ & N/A&-0.022 & -0.013 & N/A & -0.526 & -0.613\\

$7/2_1^-$ &N/A &-0.12 & -0.11 &N/A & -1.512 & -1.586 \\

\hline
$^{39}{\rm S}$   &        & &    &         \\
$3/2_1^-$ &N/A &+0.110&+0.098 &N/A & -0.820  & -0.812\\

$5/2_1^-$ &N/A &-0.058 & -0.041  &N/A & -0.680  & -0.704\\

$7/2_1^-$ &N/A &-0.094 & -0.081&N/A  & -1.127  & -1.223\\

\hline

\end{tabular}
%\end{center} 
%\end{ruledtabular}
\end{table}

%\begin{landscape}
\begin{table}
%\begin{center} 
\caption{ Comparison of calculated and experimental value of spectroscopic factors for $^{37}$S with SDPFMW (SF1) interaction. SF2 is SDPF-U calculation~\cite{Chapman} }
%\begin{ruledtabular}
\begin{tabular}{llllll}
\hline\hline
$J^\pi$ & E(SM) (keV) & \% [configuration]  &SF(exp)  & SF1 & SF2 \\
\hline
$7/2_1^-$ & 0 & 77\% $[\nu f_{7/2}^1]$ &  0.77  & 0.91  & 0.86\\

$3/2_1^-$ & 691 & 75\%$[\nu p_{3/2}^1]$   &  0.65  & 0.86  & 0.75\\

$1/2_1^-$ & 2360 & 87\% $[\nu p_{1/2}^1]$  &  0.77  & 0.97 & 0.88 \\

$5/2_1^-$ & 2660 & 79\% $[\nu (d_{3/2}s_{1/2})^{-1}\otimes \nu (f_{7/2}^{1})]$   &  0.02  & 0.008  & 0.015\\

$3/2_2^-$ & 2839 & 60\%$[\nu (d_{3/2}s_{1/2})^{-1}\otimes \nu (f_{7/2}^{1})]$   &  0.15  & 0.12 & 0.20 \\

$7/2_2^-$ & 3350 & 41\% $[\nu (d_{3/2}s_{1/2})^{-1}\otimes \nu (f_{7/2}^{1})]$   &  0.02  & 0.04 & 0.07 \\

  \hline\hline 

\end{tabular}
%\end{center} 
\end{table}
%\end{landscape}

In table 2, we have shown quadrupole and magnetic moments using $e_p$=1.5e, $e_n$=0.5e  and  $g_s^{\rm eff}$ = $g_s^{\rm free}$ for $^{35, 37,39}$S.
Our results are very close to the experimental values for $3/2_1^+$ in $^{35}$S. In the case of magnetic moment for $^{35}$S , the sign is not
 yet confirmed, our SM results predict the sign as positive. We have also predicted quadrupole and magnetic moments  for few low-lying states for
 $^{37,39}$S which are not experimentally known.  It may be useful to plan for future experiments. Finally,
for $^{37}$S presented in table 3 are the calculated and recently available experimental data~\cite{SF} on spectroscopic factors (SFs) for single nucleon transfer.
Results with SDPF-U interaction are reported in the Ref.~\cite{Chapman}. We have recalculated spectroscopic factors with SDPFMW interaction. Both calculations give reasonable agreement with the experimental data. 

\newpage 
\section{\label{sec4}Conclusions}

We have made a comprehensive shell model analysis of the latest experimental data on odd $^{35,37,39}$S
 isotopes, using SDPF-U and SDPFMW effective interactions in the $sd-pf$ valence space.

Following broad conclusions are drawn:

\begin{enumerate}
\item 
{The low energy levels are successfully reproduced, employing 0p-0h excitations for normal parity states and
 1p-1h for opposite parity states.}

\item
{Overall the SDPFMW interaction is seen to be much better for 
describing simultaneously properties of levels of both parities in
$^{37,39}$S isotopes.} 

\item
 {We have also calculated the electric quadrupole and the magnetic dipole moments for $^{35,37,39}$S
 and spectroscopic factors in $^{37}$S.  The shell model results are in
good agreement with recently available experimental data.}

\end{enumerate}

{\bf Acknowledgments: }
AS acknowledges financial support from MHRD ( Govt. of India) for her Ph.D. thesis work.
We would like to thank A. Poves for useful suggestions during this work. P.C.S. would also like to thanks B.A. Brown for his help during this work.
MJE's work is supported by Brazilian National Counsel of Technological and Scientific Development (CNPq) , Project No.165371/2015-3.

%%%%%+++++++++++++++++++++++++++++

\end{document}